\documentclass[a4paper]{PoS}

\usepackage{amsmath,xcolor,xspace}
\usepackage[utf8]{inputenc}

\definecolor{fuchsia}{RGB}{123,31,162}

\newcommand{\arxiv}{\texttt{arXiv}\xspace}
\newcommand{\eps}{\epsilon}
\newcommand{\fuchsia}{\textcolor{fuchsia}{Fuchsia}\xspace}
\newcommand{\litered}{\texttt{LiteRed}\xspace}
\newcommand{\M}[1]{\mathbb{#1}} 
\newcommand{\python}{\texttt{Python}\xspace}
\newcommand{\sage}{\texttt{SageMath}\xspace}

\title{\fuchsia and master integrals for splitting functions from differential equations in QCD}

\ShortTitle{\fuchsia and master integrals for splitting functions in QCD}

\author{\speaker{Oleksandr Gituliar}\\
        Institute of Nuclear Physics, Polish Academy of Sciences \\ ul.~Radzikowskiego 152, 31-342, Krak\'ow, Poland\\
        E-mail: \email{oleksandr.gituliar@ifj.edu.pl}}

\author{Vitaly Magerya\\
        E-mail: \email{magv@tx97.net}}

\abstract{
We report on the recent progress in reducing differential equations for Feynman master integrals to canonical form with the help of a method proposed by Roman Lee.
For the first time, we present \fuchsia\xspace --- our open-source implementation of the Lee algorithm written in \python using mathematical routines of a free computer algebra system \sage.
We demonstrate \fuchsia by reducing differential equations for NLO contributions to splitting functions in QCD, which contain both loops and legs integrals.
\\
  \begin{flushright}
    IFJPAN-IV-2016-17 \\
  \end{flushright}
}

\FullConference{Loops and Legs in Quantum Field Theory\\
		24-29 April 2016\\
		Leipzig, Germany}

\begin{document}

\section{Introduction}

More than 60 years has passed since Richard Feynman proposed a diagrammatic approach for calculating perturbative processes in quantum field theories.
Since then Feynman integrals calculus has grown to a separate branch of the mathematical physics with a big community of scientists making research in this exciting field.
With no doubt we can say that none of the recent discoveries in the high-energy particle physics could happen without precise theoretical calculations, which are based on the Feynman integrals calculation techniques. 
It is also clear that such techniques will play a key role for discoveries at the present and future high-energy colliders, hence their development and further improvement are very important task.

Recent progress in computational techniques made possible to automate calculation of loop and phase-space  Feynman integrals. Among the most popular are integration-by-parts (IBP) reduction~\cite{CT81} and the method of differential equations~\cite{Kot91a,Kot91b,Kot91c}; for a detailed overview of these and other methods see~\cite{Smi06}.

This paper is focused on the method of differential equations.
In particular, we rely on the fact that a solution to the system of DEs may be easily found as an $\eps$-series when a canonical form of this system is known~\cite{Henn13}.
We consider a general algorithm to find a canonical form of a given system of differential equations in one variable developed by Roman Lee~\cite{Lee15}.
This method describes how (1) to find a Fuchsian form of the system using modified Moser reduction algorithm~\cite{Mos59}; (2) to normalize eigenvalues of the Fuchsian system in all singular points; and if those two steps succeed,\footnote{In principle, the first step can always be done because Feynman integrals contain only logarithmic singularities.} (3) to finally transform the resulting system into canonical form.
This method is quite general, and despite our focus on systems for Feynman integrals, it can just as well be applied to reduction of any ODE system that satisfies reducibility criteria.
For alternative methods see, e.g., \cite{Henn14,ABB15}.

Since no implementation of the Lee method was publicly available so far, we close this gap by introducing \fuchsia \ --- our open-source implementation of the Lee algorithm in \python based on mathematical routines of the free computer algebra system \sage~\cite{sage}.
Combined together with Laporta algorithm \cite{Lap00} and its implementations~\cite{Smi08,MS12,Lee12,Lee13,SS13,Smi14}, these tools form a powerful tandem for evaluating multiloop and phase-space Feynman integrals.

\section{Master integrals for splitting functions}

Before we demonstrate reduction steps that \fuchsia performs to find a canonical form of a given system, let us consider next-to-leading order correction, i.e. $\mathcal{O}(\alpha_s^2)$, to the time-like splitting functions from $e^+ e^-$-annihilation in QCD.
Master integrals for real and virtual contributions to this process were calculated from difference equations in Mellin space in \cite{MM06} and from differential equations in $x$-space in \cite{Git15}.
In the next section we show how to find a canonical form of this equations with the help of \fuchsia. 
This method can be further extended for calculating next-to-next-to-leading order corrections, i.e. $\mathcal{O}(\alpha_s^3)$, to the time-like splitting functions as described in~\cite{GM15}, which are not completely know in the analytical form yet.

We start with a system of differential equations for real-virtual master integrals for NLO time-like splitting functions discussed in~\cite{Git15} which has this form:
\begin{equation}
  \frac{\partial f(x,\eps)}{\partial x} = \M M(x,\eps) f(x,\eps)
\end{equation}
where $x$ represents the fractional momentum of the final-state parton transferred to the outgoing hadron in the $e^+e^-$-annihilation process, $\eps$ is a dimensional regulator in $m=4-2\eps$ dimensions, $f(x,\eps)$ a column vector of unknown master integrals, and a coefficient matrix $\M M(x,\eps)$, generated with the help of \litered~\cite{Lee12,Lee13}, is given by
\begin{equation}
\left(
\begin{array}{cccccc}
 \frac{1-x - 2\eps (1-2x)}{(1-x)x} & 0 & 0 & 0 & 0 & 0
\\
 0 & \frac{1-x-\eps (3-4x)}{(1-x)x} & 0 & 0 & 0 & 0
\\
 0 & 0 & \frac{1-x-\eps(2-3x)}{(1-x)x} & 0 & 0 & 0
\\
 0 & -\frac{\eps(1-5\eps+6\eps^2)}{x(1-x)} & \frac{\eps(1-2\eps)^2}{(1-x)x} & \frac{2\eps}{1-x} & 0 & 0
\\
 0 & \frac{\eps(1-5\eps+6\eps^2)(3-x)}{(1-x)^2 x^3} & -\frac{2\eps(1-2\eps)^2}{(1-x)^2 x^2} & -\frac{2 \eps}{(1-x)^2 x} & -\frac{(1+2\eps)(1-2x)}{(1-x)x} & 0
\\
 0 & -\frac{2\eps(1-5\eps+6\eps^2)}{(1-x)x^2} & 0 & -\frac{4\eps}{(1-x)x} & 0 & -\frac{1+4\eps}{x} \\
\end{array}
\right)
\end{equation}%
For brevity, we do not show explicitly reduction for real-real contributions (considered in \cite{Git15} as well).
Nevertheless, by analogy with real-virtual case we provide all the corresponding results in the auxiliary files attached to this paper on \arxiv.

\section{\fuchsia and reduction to canonical form}

The whole process of reduction to canonical form is divided into three consecutive steps: fuchsification, normalization, and factorization.
Some variations to this scheme are possible, in particular to improve performance for more complicated systems; we do not consider these variations here, as they do not conceptually change the process.

\subsection{Fuchsification}

The purpose of the {\em fuchsification step} is to find an equivalent Fuchsian system together with a corresponding transformation.
(A matrix is called {\em Fuchsian} if it does not contain irregular singularities at any value of $x$, including $\infty$; in other words, Poincar\'e ranks in all singular points are zero).

Our initial matrix has three singular points, i.e., $\{0,1,\infty\}$, with Poincar\'e ranks $\{2,1,0\}$, which means that it is not in Fuchsian form.
We can find an equivalent matrix in Fuchsian form by analyzing generalized eigenvectors of residues of the initial matrix, constructing appropriate projector matrices out of their products, and performing stepwise Moser reduction with $\M P$-balance transformations between pairs of singular points.\footnote{
A $\M P$-balance between $x_1$ and $x_2$ is a basis change from $f$ to $f'$, with $f=\left(\M I - \M P + c{{x - x_2}\over{x - x_1}}\M P\right)f'$.
}
One possible equivalent Fuchsian system obtained this way looks like this:
\begin{equation}
\left(
\begin{array}{cccccc}
 \frac{2\eps}{1-x} + \frac{1-2\eps}{x} & 0 & 0 & 0 & 0 & 0
\\
 0 & \frac{\eps}{1-x} + \frac{1+3\eps}{x} & 0 & 0 & 0 & 0
\\
 0 & 0 & \frac{\eps}{1-x} + \frac{1-2\eps}{x} & 0 & 0 & 0
\\
 0 & \frac{\eps(1-2\eps)(1-3\eps)}{1-x} & -\frac{\eps(1-2\eps)^2}{1-x} & \frac{2\eps}{1-x} + \frac{1}{x} & 0 & 0
\\
 0 & \frac{\eps(1-2\eps)(1-3\eps)}{(1-x)x} & \frac{2\eps(1-2\eps)^2}{(1-x)x} & -\frac{2\eps}{(1-x)x} & \frac{2\eps}{1-x} - \frac{2\eps}{x} & 0
\\
 0 & \frac{2\eps(1-2\eps)(1-3\eps)}{(1-x)x} & 0 & -\frac{4\eps}{(1-x)x} & 0 & -\frac{4\eps}{x} \\
\end{array}
\right)
\end{equation}%


The reader can find the details of this construction in \cite{Lee15}, but it is important to note that even if the system is overall reducible to Fuchsian form, it is not always possible to construct a transformation that lowers Poincar\'e ranks at singular points without increasing Poincar\'e rank at some other points.
Sometimes the best we can do is to decrease Poincar\'e ranks in singular points at the expense of increasing them in some (arbitrary chosen) set of regular points, effectively introducing apparent singularities where there was none before.
In practice these additional apparent singularities are not a major problem, since they are subsequently removed during the normalization step.
Still, we try not to introduce them if possible to decrease intermediate expression sizes and increase overall performance.

\subsection{Normalization}

The next step is {\em normalization}.
To give you an idea of what is the aim at this stage, let us consider eigenvalues of the residues of the Fuchsian matrix we obtained in the previous step, for three singular points we have:
\begin{align*}
  x  & = 0 \quad & & \{1,\, 1-2\eps,\, 1-2\eps,\, 1-3\eps,\, -2\eps,\, -4\eps \}
  \\
  x  & = 1 \quad & & \{ -2\eps,\, -2\eps,\, -2\eps,\, -\eps,\, -\eps,\, 0 \}
  \\
  x  & = \infty \quad & & \{ 4\eps,\, 4\eps,\, -1+4\eps,\, -1+4\eps,\, -1+3\eps,\, -1+2\eps \}
\intertext{By analogy, corresponding eigenvalues for the case of real-real corrections are}
  x  & = -1 \quad & & \{ -2\eps,\, -1,\, -1,\, 1-2\eps,\, -2,\, 0,\, 0,\, 0 \}
  \\
  x  & = 0 \quad & & \{ 1-4\eps,\, 1-3\eps,\, -1-2\eps,\, -2\eps,\, 1-2\eps,\, 1-2\eps,\, 1,\, 1 \}
  \\
  x  & = 1 \quad & & \{ -2\eps,\, -2\eps,\, 1,\, 1,\, -1-2\eps,\, 1-2\eps,\, 1-2\eps,\, -1 \}
  \\
  x  & = \infty \quad & & \{ 2\eps,\, 2\eps,\, -2+3\eps,\, -2+4\eps,\, 4\eps,\, 4\eps,\, 2+4\eps,\, -2+6\eps \}
\end{align*}
All these eigenvalues are in the form $n+m\,\eps$, where $n$ is integer.
This is a commonly seen form in practical systems.
We say that matrix is normalized if $n=0$ for all eigenvalues.

In order to normalize our Fuchsian matrix we again turn to using a series of $\M P$-balances constructed out of residue eigenvectors.
This time we exploit the fact that an appropriately constructed $\M P$-balance between $x_1$ and $x_2$ will shift one of the eigenvalues at $x_1$ by $1$, and another at $x_2$ by $-1$.
Since sum of all eigenvalues is zero (because the sum of all residues is zero), by repeated application of such balances we can reach a point when our matrix is normalized at all singular points.

In the case of our example we get:
\begin{equation}
\left(
\begin{array}{cccccc}
 \frac{2\eps}{1-x} - \frac{2\eps}{x} & 0 & 0 & 0 & 0 & 0
\\
 0 & \frac{\eps}{1-x} - \frac{3\eps}{x} & 0 & 0 & 0 & 0
\\
 0 & 0 & \frac{\eps}{1-x} - \frac{2\eps}{x} & 0 & 0 & 0
\\
 0 & \frac{\eps(1-2\eps)(1-3\eps)}{1-x} & -\frac{\eps(1-2\eps)^2}{1-x} & \frac{2\eps}{1-x} & 0 & 0
\\
 0 & -\frac{\eps(1-2\eps)(1-3\eps)}{1-x} & -\frac{2\eps(1-2\eps)^2}{1-x} & \frac{2\eps}{1-x} & \frac{2\eps}{1-x} - \frac{2\eps}{x} & 0
\\
 0 & -\frac{2\eps(1-2\eps)(1-3\eps)}{1-x} & 0 & \frac{4\eps}{1-x} & 0 & -\frac{4\eps}{x} \\
\end{array}
\right)
\end{equation}%

Notice, that we rely on the fact that eigenvalues are integer in the $\eps \to 0$ limit.
This, of course, is not always the case, but sometimes it is possible to fix such a situation by a clever change of variables.
Unfortunately, we have no automated solution here; if such a need arises, users must come up with an appropriate substitution themselves.

\subsection{Factorization}

Though eigenvalues of the normalized matrix, found in the previous section, are proportional to $\eps$ the matrix itself is not.
Therefore, the purpose of the final {\em factorization} step is to put a normalized matrix to the canonical form, i.e.,
\begin{equation}
  \frac{\partial g(x,\eps)}{\partial x} = \eps \, \M S(x) \, g(x,\eps).
\end{equation}

This can normally be achieved by a constant (in $x$) transformation.

In the case of our example, the matrix $\M S(x)$ has this form
\begin{equation}
\M S(x)
=
\frac{1}{x}
\left(
\begin{array}{cccccc}
  -2 & 0 & 0 & 0 & 0 & 0
\\
 0 & -3 & 0 & 0 & 0 & 0
\\
 0 & 0 & -2 & 0 & 0 & 0
\\
 0 & 0 & 0 & 0 & 0 & 0
\\
 0 & 0 & 0 & 0 & -2 & 0
\\
 0 & 0 & 0 & 0 & 0 & -4 \\
\end{array}
\right)
+
\frac{1}{1-x}
\left(
\begin{array}{cccccc}
 2 & 0 & 0 & 0 & 0 & 0
\\
 0 & 1 & 0 & 0 & 0 & 0
\\
 0 & 0 & 1 & 0 & 0 & 0
\\
 0 & 35 & -25 & 2 & 0 & 0
\\
 0 & -35 & -50 & 2 & 2 & 0
\\
 0 & -70 & 0 & 4 & 0 & 0 \\
\end{array}
\right)
\end{equation}%
where our new basis $g(x,\eps)$ is defined as
$$f(x,\eps) = \M T(x,\eps) \, g(x,\eps)$$
and the transformation matrix $\M T(x,\eps)$ is
\begin{equation}
\M T(x,\eps) =
\left(
\begin{array}{cccccc}
 -2x & 0 & 0 & 0 & 0 & 0
\\
 0 & \frac{70x}{(1-2\eps)(1-3\eps)} & 0 & 0 & 0 & 0
\\
 0 & 0 & \frac{50x}{(1-2\eps)^2} & 0 & 0 & 0
\\
 0 & 0 & 0 & -2 & 0 & 0
\\
 -\frac{1}{(1-x)x} & -\frac{210}{(1-x)x} & 0 & 0 & \frac{2}{(1-x)x} & 0
\\
 0 & 0 & 0 & 0 & 0 & \frac{2}{x} \\
\end{array}
\right)
\end{equation}%

Note that the canonical form we have computed here is not unique, and it will be different depending on the precise sequence of reduction steps taken.
In \fuchsia we allow users to influence this process to a degree by supplying a random number generator seed; \fuchsia then selects reduction steps randomly based on that seed.
By supplying different seeds, different (but still valid) results may be obtained.

\section{Summary}

In this report we have reviewed core functionality of \fuchsia, a program for reducing differential equations for Feynman master integrals to canonical form.
\fuchsia is free and open-source: it is build in \python using a free computer algebra system \sage.
To find the corresponding analytical transformation \fuchsia uses a method proposed by Roman Lee~\cite{Lee15}, which consists of three main computational steps: fuchsification, normalization, and factorization.

In additional, an optimization for block-triangular (or sparse) matrices is also implemented, which allows \fuchsia to reduce relatively large matrices: reduction of a ${74\times74}$ matrix with 20 real and complex singular points\footnote{Provided by Costas Papadopoulos.} and at most $3\times3$ coupled blocks takes about an hour on a laptop with an Intel i5 CPU.
In spite of that we still have difficulties to reduce somewhat smaller matrices with complex singular points\footnote{Provided by Roman Lee.} due to the lack of support of factorization of polynomials with complex coefficients in \sage system, but we hope to improve this situation in the future.
Another promising direction for improvements could be support for multivariate polynomials and symbolic singular points, which would allow to reduce differential equations for multi-scale Feynman master integrals.

\section*{Acknowledgment}

We are grateful for examples of differential equation systems provided in private communication by Roman Lee and Costas Papadopoulos.

This work has been partly supported by the Polish National Center grants DEC-2013/10/E/ST2/00656 and DEC-2011/03/B/ST2/02632.
Financial support by DESY during the conference ``Loops and Legs in Quantum Field Theory 2016'' is also gratefully acknowledged.

\bibliographystyle{JHEP}
\bibliography{main}

\end{document}